\documentclass[%
reprint,
 amsmath,amssymb,
 aps,
 prx,
superscriptaddress,
]{revtex4-1}

\usepackage{indentfirst}  
\usepackage{bm}    
\usepackage{graphicx}  
\usepackage{latexsym}
\usepackage{graphicx}
\usepackage{psfrag}
\usepackage{epsfig}
\usepackage{amsmath}
\usepackage{tocloft}
\usepackage{array}
\usepackage{setspace}
\usepackage{amssymb}
\usepackage{float}
\usepackage{multirow}
\usepackage{slashbox}
\usepackage{color}
\usepackage{hyperref}
\usepackage[config=altsf]{subfig}
\usepackage{booktabs}
\usepackage{fullpage}
\usepackage[table]{xcolor}

\usepackage{graphicx}	
\usepackage{amsmath}	
\usepackage{subfig}
\usepackage{txfonts}
\usepackage{hyperref}
\usepackage{multirow}
\usepackage{xcolor}
\usepackage{caption}
\usepackage{algorithm,algorithmic,amsfonts,float}
\usepackage{url}

\begin{document}

\preprint{APS}

\title{Constraints on typical relic gravitational waves based on data of LIGO}


\author{Minghui Zhang}
\affiliation{Department of Physics, Chongqing Key Laboratory for Strongly Coupled Physics,
	Chongqing University, Chongqing 401331, China}

\author{Hao Wen}
\email[]{wenhao@cqu.edu.cn}
\affiliation{Department of Physics, Chongqing Key Laboratory for Strongly Coupled Physics,
	Chongqing University, Chongqing 401331, China}

%
%
%
%

\date{\today}

\begin{abstract}
\indent 
Relic gravitational waves (RGWs) from early universe carry very crucial and fundamental information, and thus it is of extraordinary importance to search RGW signals from data of observatories like LIGO-Virgo network. Here, focusing on typical RGWs from inflation and first-order phase transition (by sound waves and bubble collisions), effective and targeted deep learning neural networks are established to search RGW signals among real LIGO data (O2, O3a and O3b). Through lengthy adjustment and adaptation process, we construct reliable Convolutional Neural Network (CNN) to estimate the likelihood of existence  (by quantitative values and distributions) of focused RGW signals in LIGO data, or to provide constraints on their strengths. We find that if the built CNN properly estimates the parameters of RGWs, it can accurately (about $94\%$ to $99\%$) determine whether the samples contain RGW signals, and if not, the likelihood given by the CNN is not reliable. After testing large amount of LIGO datasets, the results indicate no evidence of RGWs from: 1)~inflation, 2)~sound waves, or 3)~bubble collisions, predicted by focused theories, and it provides upper limits of their GW spectral energy densities of $h^2\Omega_{gw}$ $\sim10^{-5}$, respectively for: 1)~reversely mapped 2D parameter region within rectangle of [$\beta\in(-1.85,-1.87)$ $\times$ $\alpha\in(0.005,0.007)$],   
2)~reversely mapped 3D parameter region within cuboid of [$\beta/H_{\mathrm{pt}}\in(0.02,0.16)$ $\times$ $\alpha\in(1,10)$ $\times$ $T_{\mathrm{pt}}\in(5\ast10^{9},10^{10})$ Gev], and 3)~reversely mapped 3D parameter region within cuboid of [$\beta/H_{\mathrm{pt}}\in(0.08,0.2)$ $\times$ $\alpha\in(1,10)$ $\times$ $T_{\mathrm{pt}}\in(5\ast10^{9},8\ast10^{10})$ Gev]. In short, null results and upper limits are acquire; the methods and neural networks we develop to search RGWs from LIGO data could be effective and reliable, which can be applied not only for current data but also for upcoming O4 data or other observational datasets, to establish an available scheme for exploring possible RGWs from very early stage of the cosmos or to provide constraints on relevant cosmological theories.\\

\begin{description}
\item[Keywords] relic gravitational wave, early Universe, LIGO, deep learning neural networks
\end{description}
\end{abstract}

                              
\maketitle

\section{Introduction}
In recent years, multiple GW signals from
compact binary coalescences (CBCs) have been captured by network of
LIGO-Virgo-KAGRA collaboration\cite{theligoscientificcollaboration2023gwtc3,a9,a10,akutsu2021overview}, and this groundbreaking discovery has rapidly bring us into the age of GW astronomy.
Meanwhile, the stochastic gravitational wave background (SGWB) arises as  superposition of unresolved and uncorrelated GW signals from various  origins \cite{MAGGIORE2000283,Regimbau_2011}, including astrophysical sources such as 
CBCs\cite{PhysRevD.84.084004,PhysRevD.85.104024,10.1093mnras}, neutron
stars (NSs)\cite{PhysRevD.87.063004,PhysRevD.86.104007,Ferrari}, core
collapses\cite{PhysRevD.72.084001,Howell,PhysRevD.73.104024,Zhu2010,PhysRevD.92.063005,PhysRevD.95.063015}, or cosmological sources such as  cosmic strings\cite{Kibble_1976,SARANGI2002185,PhysRevD.71.063510,PhysRevLett.98.111101,PhysRevLett.85.3761,5,9}, phase transition\cite{PhysRevD.30.272,PhysRevLett.69.2026,PhysRevD.93.104001}, inflation\cite{2,2009,PhysRevD.50.1157,PhysRevD.55.R435}, primordial black holes\cite{PhysRevLett.117.201102} and domain walls\cite{PhysRevLett.77.2879}.
The SGWB contains GWs with spectrum that spans a very wide frequency range and carries extremely rich and crucial information about fundamental problems of the universe, so this is an urgent and widely focused experimental target of GW detections. Once some components of the SGWB sit within the sensitive and frequency range of LIGO, the  recorded data  should contain signals and trace of these GWs. However, how to identify or search such signals of SGWB from these data is a crucial and challenging task, and this topic has been explored by some studies using various methods such as\cite{prl,Jiang_2023,PhysRevD.99.103015,PhysRevD.108.043025,10.1093/mnras/stac984}.\\

Particularly, the predicted relic gravitational waves (RGWs) originating from the very early stage of the universe, would be one of the most intriguing and focused targets among the SGWB. 
Detecting the RGWs is of extraordinary importance as it provides insights into the evolution of the very early universe and offers crucial evidence for various cosmological models such as the big bang theory\cite{6LK, 7LK, 8LK}.
In this article, we focus on specific components of the SGWB in origin of typical models of RGWs from inflation\cite{2005,2006,2007,2009} and the first-order phase transition (FOPT)\cite{prl,p1,p2,p3,b1,b2,p4} occurring in the early universe, and we construct effective and targeted deep learning neural networks to search signals caused by these GWs from the real data of LIGO (O2, O3a and O3b). 
Such GW targets include frequency range within the LIGO, coming with very distinctive forms of spectrum predicted by above theoretical models, and thus we try to identify such GW signals from the LIGO data according to characteristic features of these spectra. For the first step, we generate massive simulated samples of these GWs based on corresponding theoretical models, and secondly, we establish and train the   Convolutional Neural Network (CNN) by these simulated GW signals mixed with LIGO data; third, after the training, we verify and confirm that the built CNN already obtains the ability to identify these simulated GW signals according to their distinctive properties caused by various combinations of model parameters (we testify whether the trained CNN is able to estimate the model parameters of large number of different signal samples to assess the performance and reliability of the pipeline); the next, we search above GW signals by such verified neural network among the real LIGO data, for different orders of magnitude of energy density or amplitude of these typical GWs, to find out the likelihood that the data contain above GW signals in such levels, and if there is no evidence for these signals, we  can provide constraints or upper limits on corresponding strengths of these GWs.\\

The theoretical models for inflation stage we focus in this article predict the RGWs in a wide frequency band of about $10^{-18}$ to $10^{10}$Hz\cite{2005,2006,2007}, and in some parameter range, their amplitude could sit within the sensitivity level of the LIGO. 
The model of FOPT\cite{prl} we focus in this article is anticipated to generate components of SGWB through processes involving bubble nucleation, expansion, collision, and thermalization into light particles\cite{prl,b1}.
In this model, if the parameter $T_{pt}\sim(10^{7}-10^{10})$ Gev, the resulting SGWB falls within the frequency range of Advanced LIGO and Advanced Virgo\cite{41,42}. During the FOPT, it has been determined that the GW can be mainly generated from three sources: bubble collisions, sound waves, and magnetohydrodynamic turbulence\cite{88,p4,50,51}. 
Here, we do not consider  contribution of the magnetohydrodynamic turbulence, because it always occurs with the sound waves and its strength is secondary, and furthermore, we notice that its spectrum is the least known\cite{50,53,54,55,56}. Thus, we mainly consider  the GWs generated from the sound waves and bubble collisions in the FOPT.\\

Plan of this article is as follows: in Sect. \ref{sec2}, we generate simulated GW signals based on the focused theoretical models, and we mixed them together with the real LIGO data, to prepare the datasets for the training and testing for the constructed deep learning neural networks. In Sect. \ref{sec3}, we try to identify above GW signals from the real LIGO data or estimate the likelihood that such signals are contained, and based on it the constraints of their strengths are given. In Sect. \ref{sec4}, we provide a conclusion and discuss the findings of our research.\\
~\\

\section{Generation of samples by simulated Gravitational wave datasets and mixed with real LIGO data}\label{sec2}

The RGWs could be generated during the inflationary expansion of the early universe and have a spectrum distributed over a very wide range of frequencies. The typical primordial spectrum of such RGWs from inflation (we abbreviate this source as RGWinfl, the same hereafter) can be expressed as\cite{2009}:
\begin{equation}\label{eq1}
	h\left(\nu, \tau_{i}\right)=A\left(\frac{\nu}{\nu_{H}}\right)^{2+\beta} A_{\alpha_{t}}(\nu)
\end{equation}

For spectra in other frequency ranges, the calculation is the same as \cite{2005, 2006, 2007}. In the frequency range of 20-300Hz, the RGWs can be described as\cite{2006,2009}:
\begin{equation}\label{eq2}
	\begin{aligned}
		h\left(\nu, \tau_{H}\right) \approx A\left(\frac{\nu}{\nu_{H}}\right)^{\beta+1} \frac{\nu_{H}}{\nu_{2}} \frac{1}{\left(1+z_{E}\right)^{3+\epsilon}}A_{\alpha_{t}}(\nu)
	\end{aligned}
\end{equation}

The inflationary index $\beta$ is an important model parameter that affects the overall slope of the RGW spectrum. Where the extra factor\cite{2009}
\begin{equation}\label{eq3}
	A_{\alpha_{t}}(\nu) \equiv\left(\frac{\nu}{\nu_{0}}\right)^{(1 / 4) \alpha_{t} \ln \left(\nu / \nu_{0}\right)}
\end{equation}

is the deviation from a simple power-law spectrum caused by $\alpha_t$, which reflects the additional curvature. The dimensionless spectral energy density of RGWs can be described as\cite{2009}:
\begin{equation}\label{eqreg}
	\Omega_{g}(\nu)=\frac{\pi^{2}}{3} h^{2}\left(\nu, \tau_{H}\right)\left(\frac{\nu}{\nu_{H}}\right)^{2}
\end{equation}

In Eq. \ref{eq2}, the $A=4.94 \times 10^{-5} r^{1 / 2}\left({\nu_{H}}/{\nu_{0}}\right)^{2+\beta}$; the factor $A$ contains some oscillating factors in the form of $\cos(k{\tau_H})$ or $\cos(y_2)$\cite{2006}. The small parameter $\epsilon\equiv(1+\beta)(1-\gamma) / \gamma$, $1+z_{E}=\left({\Omega_{\Lambda}}/{\Omega_{m}}\right)^{1 / 3}$, $\Omega_{\Lambda}$ is for dark energy, $\Omega_{m}$ is for dark matter. In this article, we take $\gamma=1.044$, $\Omega_{\Lambda}=0.75$, $\Omega_{m}=0.25$, tensor/scalar ratio $r=0.55$.
The energy density spectrum of the SGWB is: $\Omega_{\mathrm{GW}}(f)=d \rho_{\mathrm{GW}} /\left(\rho_{c} d \ln f\right)$, $\rho_{c}$ is the current critical energy density and $\rho_{c}=3 c^{2} H_{0}^{2} /(8 \pi G)$. In the most common thermal transitions in the early universe, the main source of GW production is the sound waves in the plasma caused by the coupling between the scalar field and the thermal bath\cite{prl,p1,p2,p3}. The spectrum from numerical simulations can be expressed as\cite{prl}:
\begin{equation}\label{eq4}
	\begin{aligned}\Omega_{\mathrm{sw}}(f) h^{2}= & 2.65 \times 10^{-6}\left(\frac{H_{\mathrm{pt}}}{\beta}\right)\left(\frac{\kappa_{\mathrm{sw}} \alpha}{1+\alpha}\right)^{2}\left(\frac{100}{g_{*}}\right)^{1 / 3} \\& \times v_{w}\left(\frac{f}{f_{\mathrm{sw}}}\right)^{3}\left(\frac{7}{4+3\left(f / f_{\mathrm{sw}}\right)^{2}}\right)^{7 / 2} \Upsilon\left(\tau_{\mathrm{sw}}\right)\end{aligned}
\end{equation}
where $\kappa_\mathrm{sw}$ is the fraction of vacuum energy converted into bulk fluid kinetic energy; $H_{\mathrm{pt}}$ is the Hubble parameter at the temperature $T_{\mathrm{pt}}$; $g_*$ is the number of relativistic degrees of freedom, which we take to be 100 in this paper; $f_\mathrm{sw}$ is the present peak frequency\cite{prl},
\begin{equation}\label{eq5}
	f_{\mathrm{sw}}=19 \frac{1}{v_{w}}\left(\frac{\beta}{H_{\mathrm{pt}}}\right)\left(\frac{T_{\mathrm{pt}}}{100 \mathrm{GeV}}\right)\left(\frac{g_{*}}{100}\right)^{1 / 6} \mu \mathrm{Hz}
\end{equation}
\begin{equation}\label{eq6}
	\Upsilon=1-\left(1+2 \tau_{\mathrm{sw}} H_{\mathrm{pt}}\right)^{-1 / 2}
\end{equation}
$\tau_{\mathrm{sw}}$ is usually chosen as the time scale of the onset of turbulence\cite{p4}, $\tau_{\mathrm{sw}} \approx R_{\mathrm{pt}} / \bar{U}_{f}$, the exponential nucleation of bubbles $R_{\mathrm{pt}}=(8 \pi)^{1 / 3} v_{w} / \beta$, and $\bar{U}_{f}^{2}=3 \kappa_{\mathrm{sw}} \alpha /[4(1+\alpha)]$\cite{p4}. 
When sound waves and magnetohydrodynamic turbulence are highly suppressed or absent, bubble collisions may dominate, and the resulting GW spectrum can be well modeled by the envelope approximation. The spectrum is\cite{prl,b1,b2}:
\begin{equation}\label{eq8}
	\begin{aligned}\Omega_{\text {coll }}(f) h^{2}= & 1.67 \times 10^{-5} \Delta\left(\frac{H_{\mathrm{pt}}}{\beta}\right)^{2}\left(\frac{\kappa_{\phi} \alpha}{1+\alpha}\right)^{2} \\& \times\left(\frac{100}{g_{*}}\right)^{1 / 3} S_{\text {env }}(f)\end{aligned}
\end{equation}
where amplitude $\Delta\left(v_{w}\right)=0.48 v_{w}^{3} /\left(1+5.3 v_{w}^{2}+5 v_{w}^{4}\right)$. The shape of the spectrum is
\begin{equation}\label{eq9}
	S_{\mathrm{env}}=1 /\left(c_{l} \tilde{f}^{-3}+\left(1-c_{l}-c_{h}\right) \tilde{f}^{-1}+c_{h} \tilde{f}\right)
\end{equation}
where $c_{l}=0.064$, $c_{h}=0.48$, $\tilde{f}=f / f_{\mathrm{env}}$, $f_{\mathrm{env}}$ presents the current peak frequency
\begin{equation}\label{eq10}
	f_{\mathrm{env}}=16.5\left(\frac{f_{\mathrm{bc}}}{\beta}\right)\left(\frac{\beta}{H_{\mathrm{pt}}}\right)\left(\frac{T_{\mathrm{pt}}}{100 \mathrm{GeV}}\right)\left(\frac{g_{*}}{100}\right)^{1 / 6} \mu \mathrm{Hz}
\end{equation}
\begin{equation}\label{eq11}
	f_{\mathrm{bc}}=0.35 \beta /\left(1+0.069 v_{w}+0.69 v_{w}^{4}\right)
\end{equation}
The characteristic amplitude of GWs can be expressed as\cite{wz}:
\begin{equation}\label{eq12}
	h_{\text {c}}(f) \simeq 1.263 \times 10^{-18}(1 \mathrm{~Hz} / f) \sqrt{h^{2} \Omega_{\text {gw }}(f)}
\end{equation}
In our analysis, we take $v_{w}=1$ and $\kappa_{\phi}=1$, where $\kappa_{s w}=\frac{\alpha}{0.73+0.083 \sqrt{\alpha}+\alpha}$\cite{prl,p4}.\\

\begin{figure}
	\centering
	\subfloat[\centering]{{\includegraphics[height = 3.3cm, width = 3.5cm]{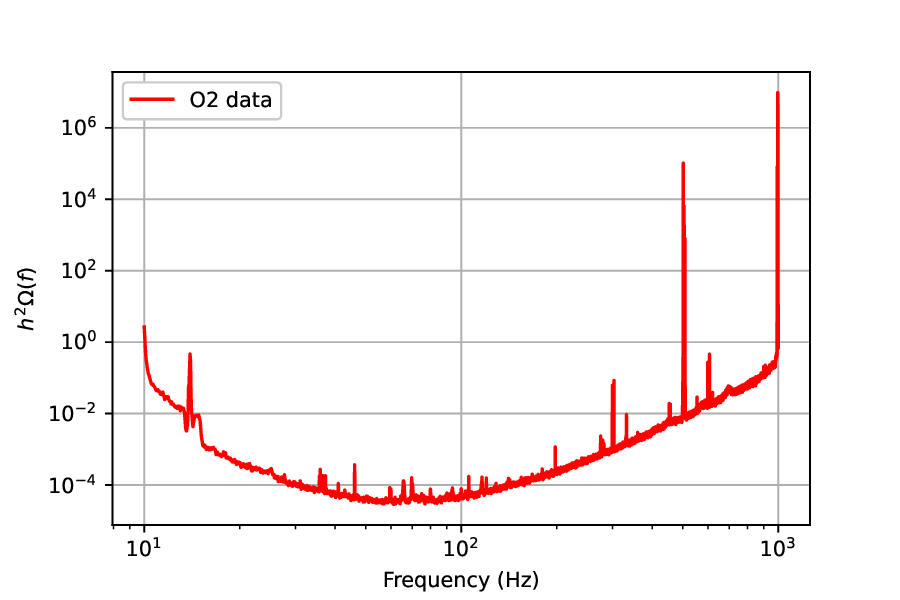} }}%
	\qquad
	\subfloat[\centering]{{\includegraphics[height = 3.3cm, width = 3.5cm]{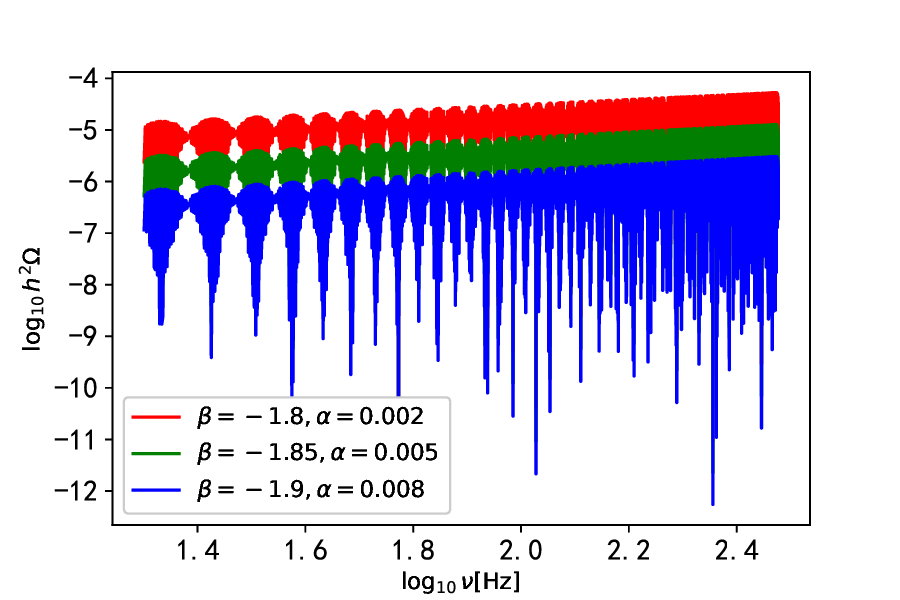} }}%
	\qquad
	\subfloat[\centering]{{\includegraphics[height = 3.3cm, width = 3.5cm]{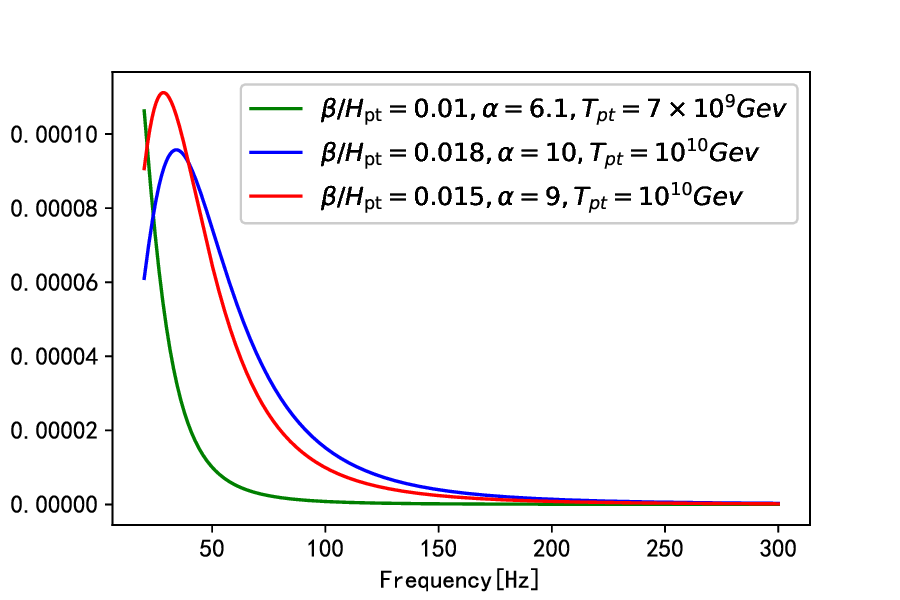} }}%
	\qquad
	\subfloat[\centering]{{\includegraphics[height = 3.3cm, width = 3.5cm]{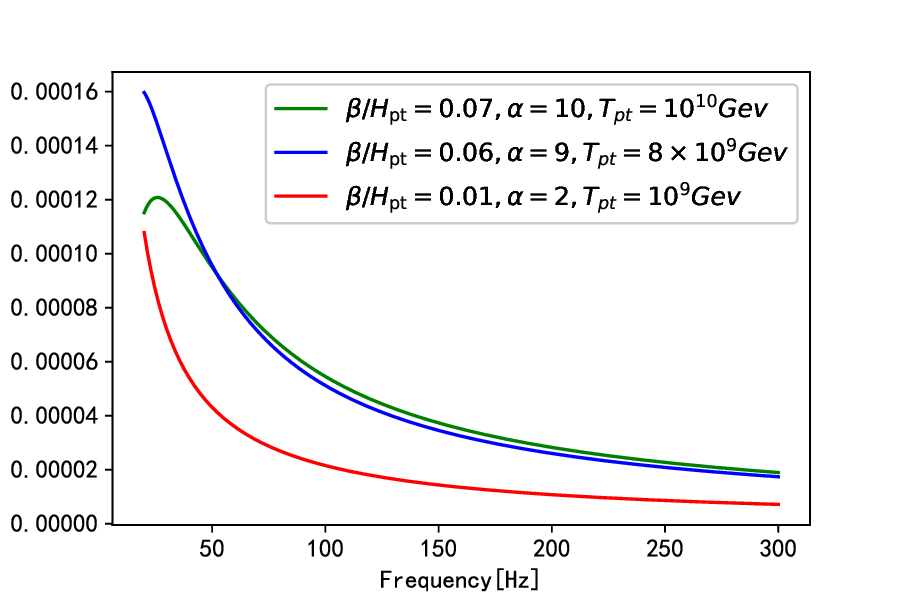} }}%
	\caption{Subfigure (a) presents the spectral energy density ($h^2\Omega$) of typical O2 data; subfigure (b) presents typical spectral energy density ($h^2\Omega$) of RGWinfl; subfigure (c) demonstrates the spectral energy density ($h^2\Omega$) of GWs originating from sound waves in several parameters; subfigure (d) demonstrates some example spectral energy density ($h^2\Omega$) of GWs originating from bubble collisions.}%
	\label{fig1}%
\end{figure}

\begin{figure}
	\centering
	\subfloat[\centering]{{\includegraphics[height = 3cm, width = 3.5cm]{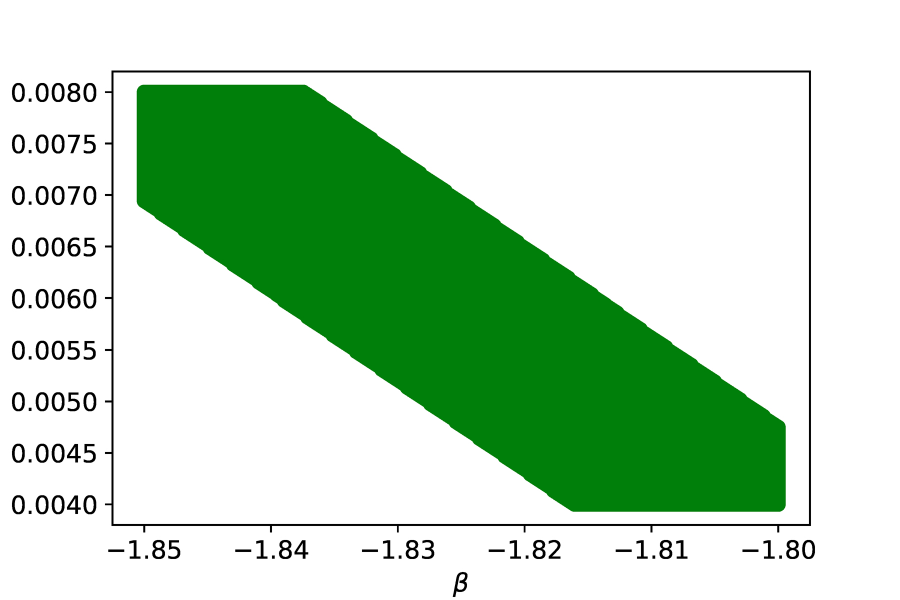} }}%
	\qquad
	\subfloat[\centering]{{\includegraphics[height = 3cm, width = 3.5cm]{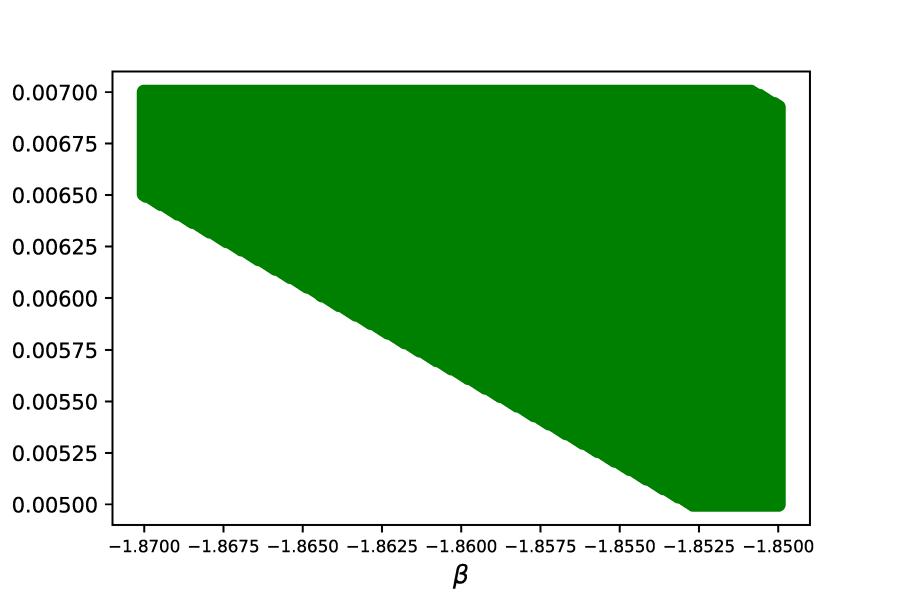} }}%
	\qquad
	\subfloat[\centering]{{\includegraphics[height = 3cm, width = 3.5cm]{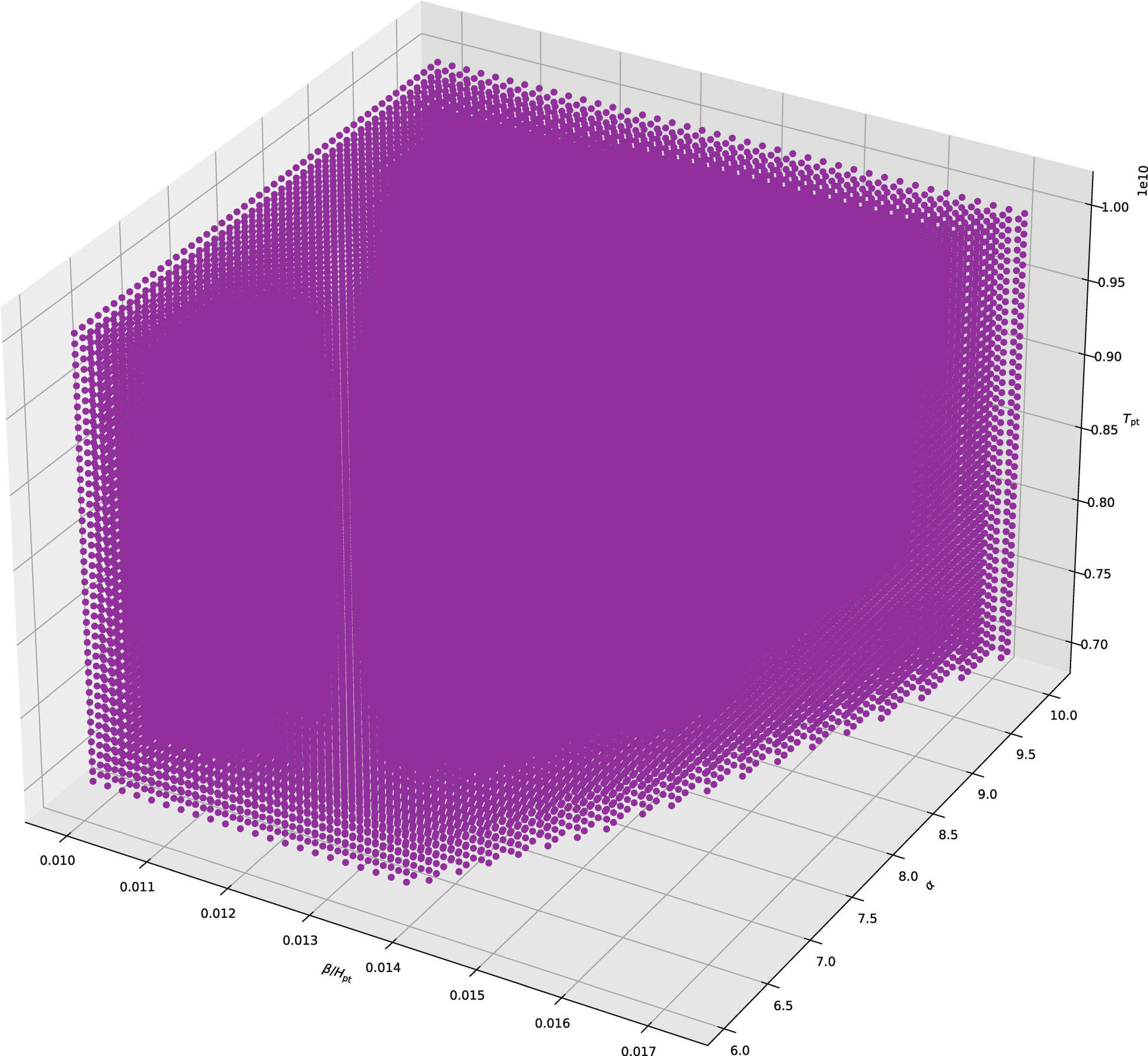} }}%
	\qquad
	\subfloat[\centering]{{\includegraphics[height = 3cm, width = 3.5cm]{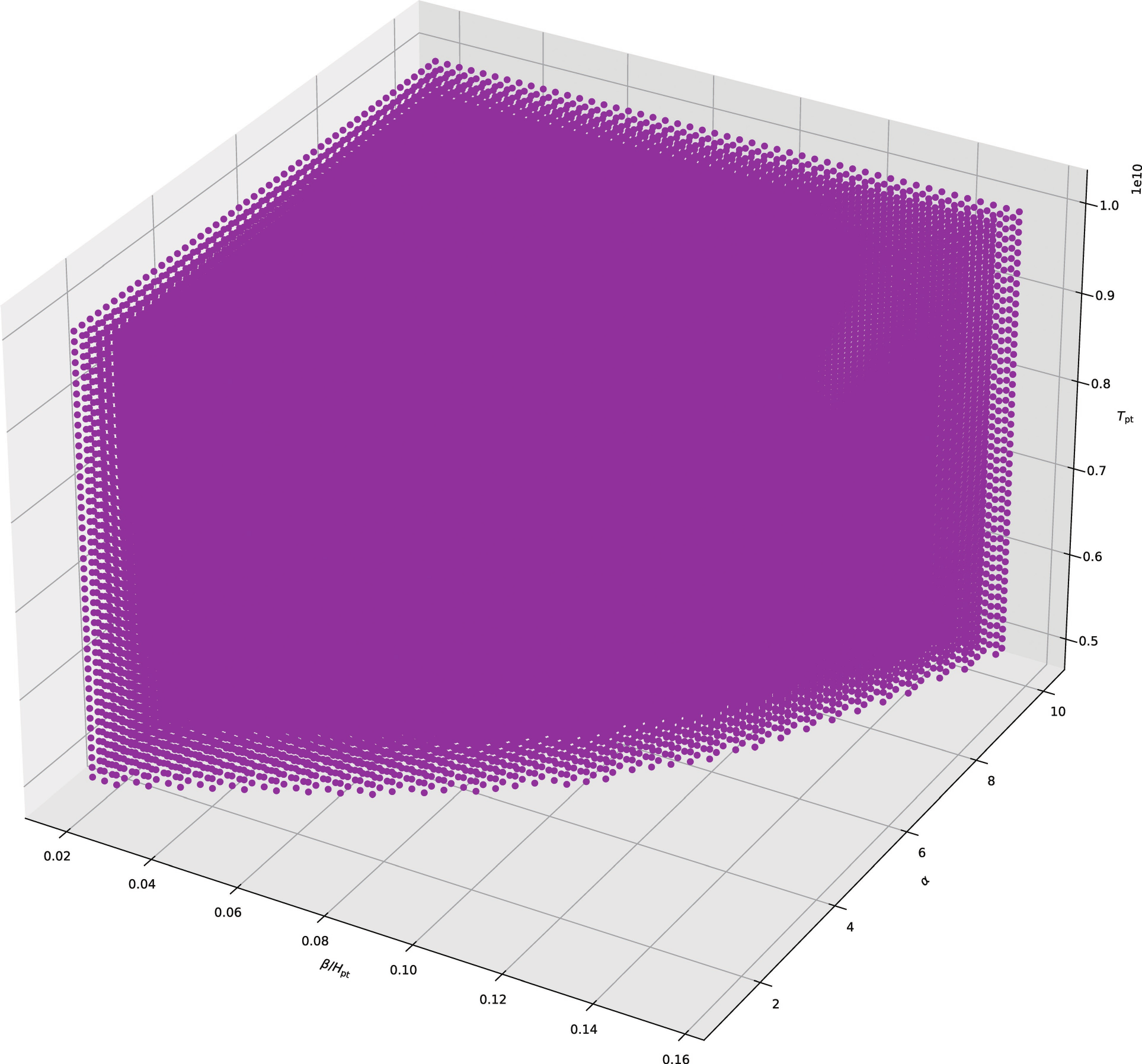} }}%
	\qquad
	\subfloat[\centering]{{\includegraphics[height = 3cm, width = 3.5cm]{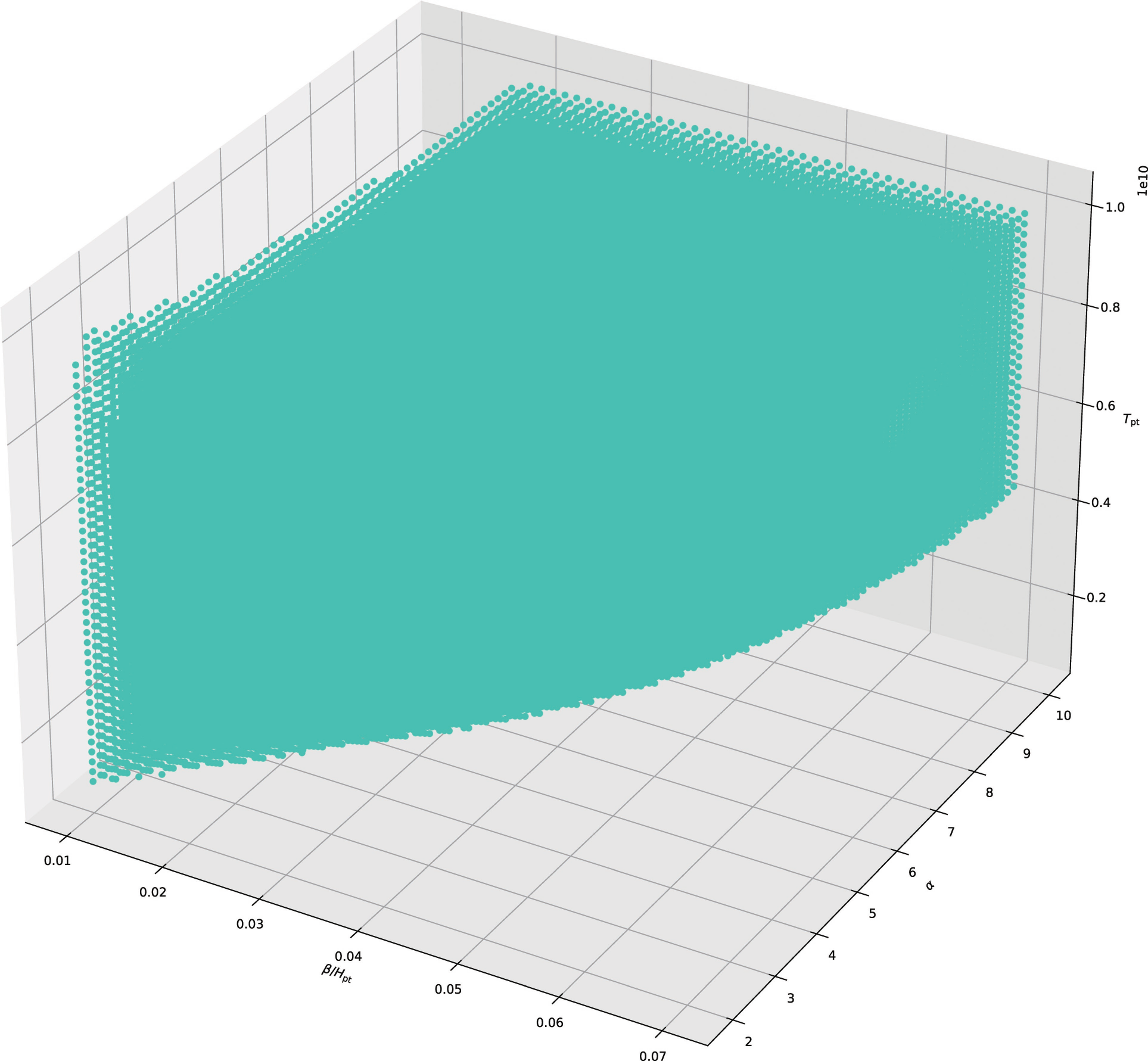} }}%
	\qquad
	\subfloat[\centering]{{\includegraphics[height = 3cm, width = 3.5cm]{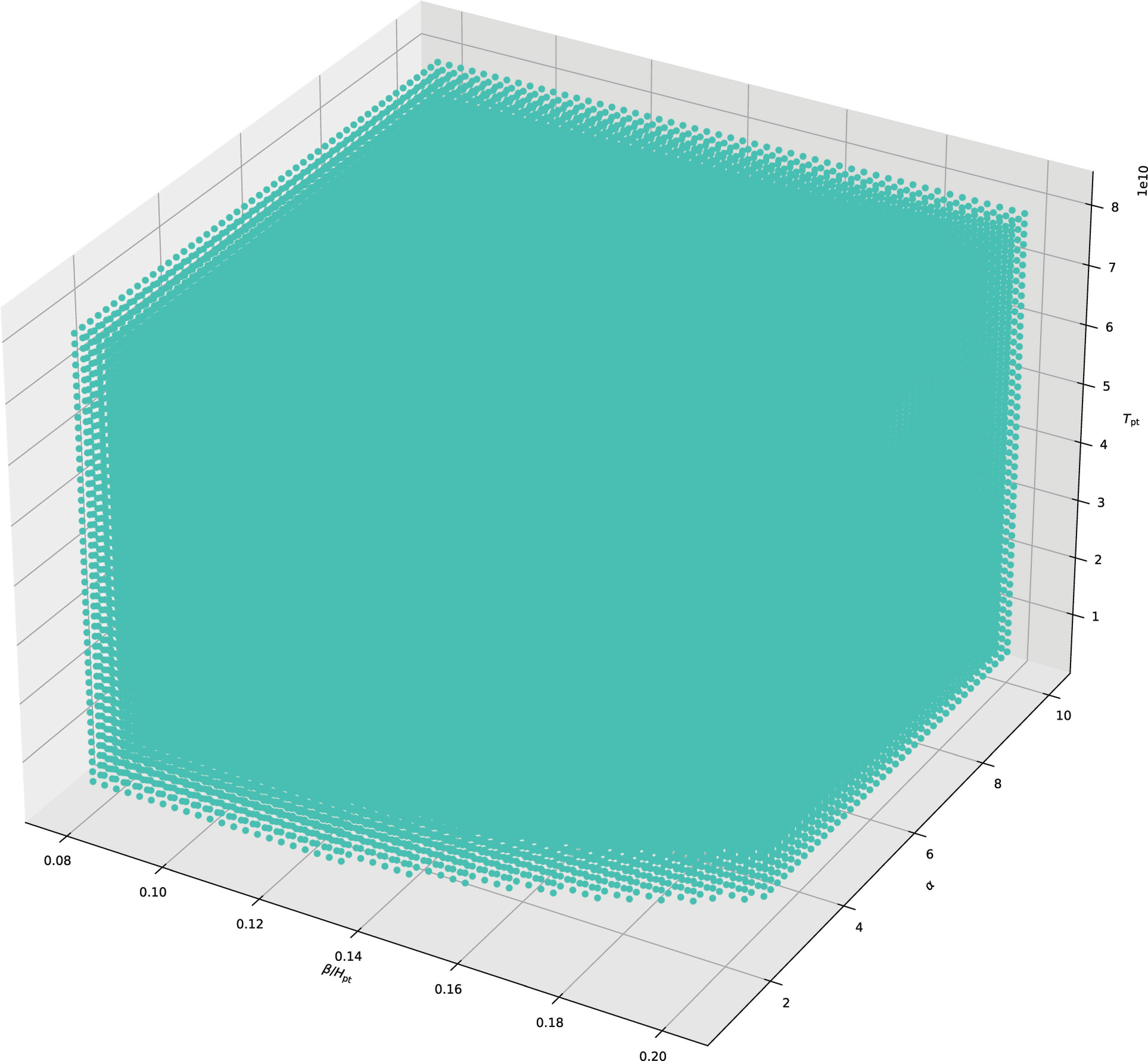} }}%
	\caption{By reverse mapping of parameters of RGW models, subfigures (a) and (b) show the parameter space for the sets 1 and 2 in Table \ref{tab2}, respectively. Subfigures (c), (d), (e) and (f) display the parameter space for the sets 1, 2, 4 and 5 in Table \ref{tab3}, respectively. }%
	\label{figcskj}%
\end{figure}

We obtain O2 data from the Gravitational Wave Open Science Center (GWOSC)\cite{gwo,J,wh}. The selected segments are from real strain data, and each dataset contains 1-second time series, with a sampling rate of 4096Hz. We use PyCBC to filter the datasets, making their frequencies between 20-300Hz, and transform them into the frequency domain. Then we take the modulus of the frequency domain data and divide it by $\sqrt{2}$ to get the characteristic amplitude. The spectral energy density of O2 data can be acquired from Eq\ref{eq12}.  The O2 data samples we obtained are twice the number of simulated GW signal samples (as described below). We consider three GW models of Eqs. (\ref{eqreg}), (\ref{eq4}) and (\ref{eq8}), respectively. In fact, although deep neural networks have the ability to generalize, there is still a possibility of mis-classification, especially when there is a large difference in the parameter space.  Therefore, in order to ensure the reliability and robustness of the applied method and neural network, we use reverse mapping to acquire the specific shape (usually not rectangle or cuboid, see Fig. \ref{figcskj}) of the parameter regions   which particularly correspond to different order of magnitude of GW strength, and thus generate samples within these specific parameter 2D or 3D regions.  Concretely, for case of RGWinfl, for each order of magnitude of spectral energy density $h^2\Omega_{GW}$, we take $300\times300$ points sitting in the above mentioned corresponding 2D parameter surface within the rectangle of $\beta$ $\times$  $\alpha$ values (see table \ref{tab2}). For GWs from FOPT, for each order of magnitude of spectral energy density $h^2\Omega_{GW}$, we take $50\times50\times45$ points in the 3D parameter volume within the cuboid of $\beta/H_{\mathrm{pt}}$ $\times$ $\alpha$ $\times$ $T_{\mathrm{pt}}$ (see table \ref{tab3}).
Namely, the actually parameter regions we selected are within and smaller than the ``parameter boundary'' of Table \ref{tab2} and \ref{tab3}, so they more accurately cover the corresponding parameter points and avoid mismatched parameter selection.
In this way, the selected parameter points fitly cover every corresponding order of magnitude of GWs. With generating simulated samples of above GWs in forms of spectral energy densities, we overlay them with spectral energy densities of O2 data. Among all the obtained samples, in which half are pure O2 data samples and the other half are samples of O2 data plus the simulated GW signals. We split the training set and test set in a 7:3 ratio.

\section{Estimation of likelihood of targeted GW signals in real LIGO data by using deep learning neural networks}\label{sec3}

We construct a suitable one-dimensional CNN model after lengthy adjustment and adaptation process, using the alternation of convolutional layers and pooling layers to extract the features of the samples. The structure of the CNN model is shown in Table \ref{tab1}.

\begin{table} 
	\centering
	\caption{The structure of CNN used to calculate the confidence of GW signals in the real LIGO data.}
	\label{tab1}
	\begin{tabular}{c|c|c} 
		\hline  
		Layer type & Channel & Kernel size \\ 
		\hline
		Input &  &   \\
		Conv1D+Relu & 8 & 4 \\
		MaxPool1D &  & 4 \\
		Conv1D+Relu & 16 & 8 \\
		MaxPool1D &  & 3 \\
		Conv1D+Relu & 32 & 3 \\
		MaxPool1D &  & 2 \\
		Flatten &  &  \\
		Dropout(0.8) &  &  \\
		Dense+Softmax(2) &  &  \\
		\hline
	\end{tabular}
\end{table}
Before training, we normalize the sample data to between 0 and 1. Using relu\cite{relu} as the activation function, using the Categorical Cross-Entropy(CCE) loss function to evaluate the deviation between the predicted values and the actual values, using the Adam optimizer\cite{adam1,adam2,YHQ} to optimize the weights and biases of the CNN, and setting the learning rate to $10^{-5}$. Finally, we employ the binary output score $s$ of the network to calculate the confidence of the GW signal in LIGO data through the softmax function\cite{wh}:
\begin{equation}\label{eqs}
	p=\frac{1}{1+e^{-s}}
\end{equation}
The $p$-value is in the range of 0-1, and we adopt it to characterize the confidence of that the LIGO data contain typical RGW signals. In this paper, we use 0.5 as the threshold, and closer the $p$-value is to 1, the higher the likelihood that the data contain the targeted GW signals. In our analysis, we first generate simulated GW signals of different orders of magnitudes using different parameter spaces (as mentioned in the Sect. \ref{sec2}), then we train CNNs for signals of different orders of magnitudes separately. In each parameter space for trained CNN model, we calculate the confidence that the real LIGO data (based on the 6000 raw sample obtained by O2, O3a and O3b data, respectively) contain the above mentioned three types of potential RGW signals. We give the distribution and mean values of the confidence that the O2, O3a and O3b data would contain these GW signals.
Table \ref{tab2} and \ref{tab3} respectively shows the parameter space of three GW models that we consider, and corresponding mean values of the confidence or likelihood.
\begin{table*}
	\centering
	\caption{Parameter settings for RGWinfl and the mean value ($p$-value) of the confidence of the real LIGO data (O2, O3a and O3b) containing such GW signals. The ``Recognition accuracy'' is for the constructed CNN in determining whether the test data is mixed with the simulated GW signals, in addition to the real LIGO data.}
	\label{tab2}
	\begin{tabular}{c|c|c|c|c|c|c|c|c}
		\hline
		\hline
		\multirow{5}{0.3cm}{set} & \multirow{5}{2.3cm}{Parameter boundary} & \multirow{5}{1.5cm}{magnitude of $h^2\Omega_{GW}$} & \multirow{5}{1cm}{$\sqrt{PSD}$ ($Hz^{-1/2}$)} & \multirow{5}{1.8cm}{Recognition accuracy} & \multirow{5}{2.3cm}{mean $p$-value for O2 data plus simulated GW signals} & \multirow{5}{1.3cm}{mean $p$-value (O2)} & \multirow{5}{1.3cm}{mean $p$-value (O3a)} & \multirow{5}{1.3cm}{mean $p$-value (O3b)} \\
		~ & ~ & ~ & ~ & ~ & ~ & ~ & ~ & ~ \\
		~ & ~ & ~ & ~ & ~ & ~ & ~ & ~ & ~ \\
		~ & ~ & ~ & ~ & ~ & ~ & ~ & ~ & ~ \\
		~ & ~ & ~ & ~ & ~ & ~ & ~ & ~ & ~ \\
		\hline
		\multirow{2}*{1} & $\beta\in(-1.8,-1.85)$ & \multirow{2}*{$\sim10^{-4}$} & \multirow{2}*{$\sim10^{-22}$} & \multirow{2}*{0.9982} & \multirow{2}*{0.99986713} & \multirow{2}*{0.00776} & \multirow{2}*{$3.15\times10^{-7}$} & \multirow{2}*{$3.37\times10^{-5}$}\\
		~ & $\alpha\in(0.004,0.008)$ & ~ & ~ & ~ & ~ & ~ & ~ & ~ \\
		\hline
		\multirow{2}*{2} & $\beta\in(-1.85,-1.87)$ & \multirow{2}*{$\sim10^{-5}$} & \multirow{2}*{$\sim10^{-23}$} & \multirow{2}*{0.9422} & \multirow{2}*{0.9033364} & \multirow{2}*{0.10817} & \multirow{2}*{$0.0035649$} & \multirow{2}*{$0.0005467$}\\
		~ & $\alpha\in(0.005,0.007)$ & ~ & ~ & ~ & ~ & ~ & ~ & ~ \\
		\hline
		\multirow{2}*{3} & $\beta\in(-1.88,-1.91)$ & \multirow{2}*{$\sim10^{-6}$} & \multirow{2}*{$\sim10^{-24}$} & \multirow{2}*{0.4961} & \multirow{2}*{0.501124
		} & \multirow{2}*{0.498391} & \multirow{2}*{0.4884417} & \multirow{2}*{0.487882}\\
		~ & $\alpha\in(0.006,0.008)$ & ~ & ~ & (Unrecognized) & ~ & ~ & ~ & ~ \\
		\hline
		\multirow{2}*{4} & $\beta\in(-1.92,-2.0)$ & $\sim10^{-7}$ to & $\sim10^{-25}$ to & {0.4917} & {0.5022312} & {0.500165} & {0.498763} & {0.489765}\\
		~ & $\alpha\in(0.007,0.008)$ & $\sim10^{-10}$ & $\sim10^{-28}$ & (Unrecognized)   & ~ & ~ & ~ & ~ \\
		\hline
	\end{tabular}
\end{table*}

\begin{table*}
	\centering
	\caption{Parameter settings for GWs from FOPT and the mean $p$-value of the confidence of the real LIGO data (O2, O3a and O3b) containing these GW signals.}
	\label{tab3}
	\begin{tabular}{c|c|c|c|c|c|c|c|c}
		\hline
		\hline
		\multicolumn {9} {c} {GWs originating from sound waves} \\
		\hline
		\multirow{5}{0.3cm}{set} & \multirow{5}{2.5cm}{Parameter boundary} & \multirow{5}{1.4cm}{magnitude of $h^2\Omega_{GW}$} & \multirow{5}{1cm}{$\sqrt{PSD}$ ($Hz^{-1/2}$)} & \multirow{5}{1.3cm}{Recognition accuracy} & \multirow{5}{2.3cm}{mean  $p$-value  for O2 data plus simulated GW signals} & \multirow{5}{1cm}{mean $p$-value (O2)} & \multirow{5}{1.3cm}{mean $p$-value (O3a)} & \multirow{5}{1.3cm}{mean $p$-value (O3b)} \\
		~ & ~ & ~ & ~ & ~ & ~ & ~ & ~ & ~ \\
		~ & ~ & ~ & ~ & ~ & ~ & ~ & ~ & ~ \\
		~ & ~ & ~ & ~ & ~ & ~ & ~ & ~ & ~ \\
		~ & ~ & ~ & ~ & ~ & ~ & ~ & ~ & ~ \\
		\hline
		\multirow{3}*{1} & $\beta/H_{\mathrm{pt}}\in(0.01,0.019)$ & \multirow{3}*{$\sim10^{-4}$} & \multirow{3}*{$\sim10^{-22}$} & \multirow{3}*{0.9994} & \multirow{3}*{0.994755} & \multirow{3}*{0.02574} & \multirow{3}*{$4\times10^{-6}$} & \multirow{3}*{$1\times10^{-4}$}\\
		~ & $\alpha\in(6.1,10)$ & ~ & ~ & ~ & ~ & ~ & ~ & ~ \\
		~ & $T_{\mathrm{pt}}\in(7\times10^{9},10^{10})$Gev & ~ & ~ & ~ & ~ & ~ & ~ & ~ \\
		\hline
		\multirow{3}*{2} & $\beta/H_{\mathrm{pt}}\in(0.02,0.16)$ & \multirow{3}*{$\sim10^{-5}$} & \multirow{3}*{$\sim10^{-23}$} & \multirow{3}*{0.9494} & \multirow{3}*{0.94873} & \multirow{3}*{0.12302} & \multirow{3}*{0.00151} & \multirow{3}*{0.00058}\\
		~ & $\alpha\in(1,10)$ & ~ & ~ & ~ & ~ & ~ & ~ & ~ \\
		~ & $T_{\mathrm{pt}}\in(5\times10^{9},10^{10})$Gev & ~ & ~ & ~ & ~ & ~ & ~ & ~ \\
		\hline
		\multirow{3}*{3} & $\beta/H_{\mathrm{pt}}\in(0.17,0.4)$ & \multirow{3}*{$\sim10^{-6}$} & \multirow{3}*{$\sim10^{-24}$} & {0.5012} & \multirow{3}*{0.499987} & \multirow{3}*{0.49341} & \multirow{3}*{0.48976} & \multirow{3}*{0.48875}\\
		~ & $\alpha\in(1.1,10)$ & ~ & ~ & (Unrecognized) & ~ & ~ & ~ & ~ \\
		~ & $T_{\mathrm{pt}}\in(5\times10^{8},10^{10})$Gev & ~ & ~ & ~ & ~ & ~ & ~ & ~ \\
		\hline
		\hline
		\multicolumn {9} {c} {GWs originating from bubble collisions}\\
		\hline
		\multirow{3}*{4} & $\beta/H_{\mathrm{pt}}\in(0.01,0.07)$ & \multirow{3}*{$\sim10^{-4}$} & \multirow{3}*{$\sim10^{-22}$} & \multirow{3}*{0.9998} & \multirow{3}*{0.999396} & \multirow{3}*{0.002406} & \multirow{3}*{$6\times10^{-5}$} & \multirow{3}*{$1\times10^{-5}$}\\
		~ & $\alpha\in(2,10)$ & ~ & ~ & ~ & ~ & ~ & ~ & ~ \\
		~ & $T_{\mathrm{pt}}\in(10^9,10^{10})$Gev & ~ & ~ & ~ & ~ & ~ & ~ & ~ \\
		\hline
		\multirow{3}*{5} & $\beta/H_{\mathrm{pt}}\in(0.08,0.2)$ & \multirow{3}*{$\sim10^{-5}$} & \multirow{3}*{$\sim10^{-23}$} & \multirow{3}*{0.9683} & \multirow{3}*{0.968897} & \multirow{3}*{0.062567} & \multirow{3}*{0.002691} & \multirow{3}*{0.001658}\\
		~ & $\alpha\in(1,10)$ & ~ & ~ & ~ & ~ & ~ & ~ & ~ \\
		~ & $T_{\mathrm{pt}}\in(5\times10^{9},8\times10^{10})$Gev & ~ & ~ & ~ & ~ & ~ & ~ & ~ \\
		\hline
		\multirow{3}*{6} & $\beta/H_{\mathrm{pt}}\in(0.3,0.7)$ & \multirow{3}*{$\sim10^{-6}$} & \multirow{3}*{$\sim10^{-24}$} & {0.4992} & \multirow{3}*{0.50231} & \multirow{3}*{0.49654} & \multirow{3}*{0.49002} & \multirow{3}*{0.48532}\\
		~ & $\alpha\in(4,10)$ & ~ & ~ & (Unrecognized) & ~ & ~ & ~ & ~ \\
		~ & $T_{\mathrm{pt}}\in(2\times10^{8},10^{10})$Gev & ~ & ~ & ~ & ~ & ~ & ~ & ~ \\
		\hline
	\end{tabular}
\end{table*}

\begin{figure*}
	\centering
	\subfloat[\centering]{{\includegraphics[height = 4cm, width = 6cm]{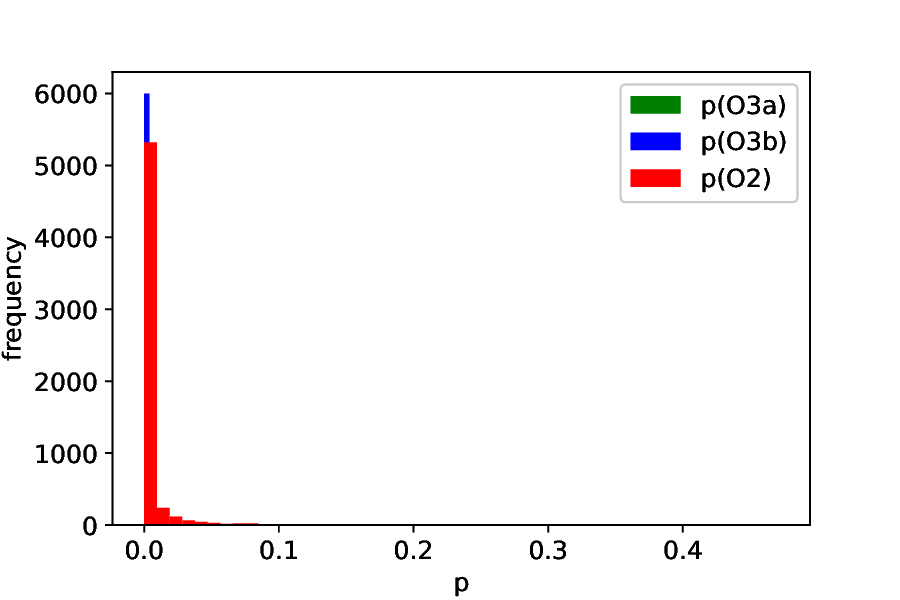} }}%
	\qquad
	\subfloat[\centering]{{\includegraphics[height = 4cm, width = 6cm]{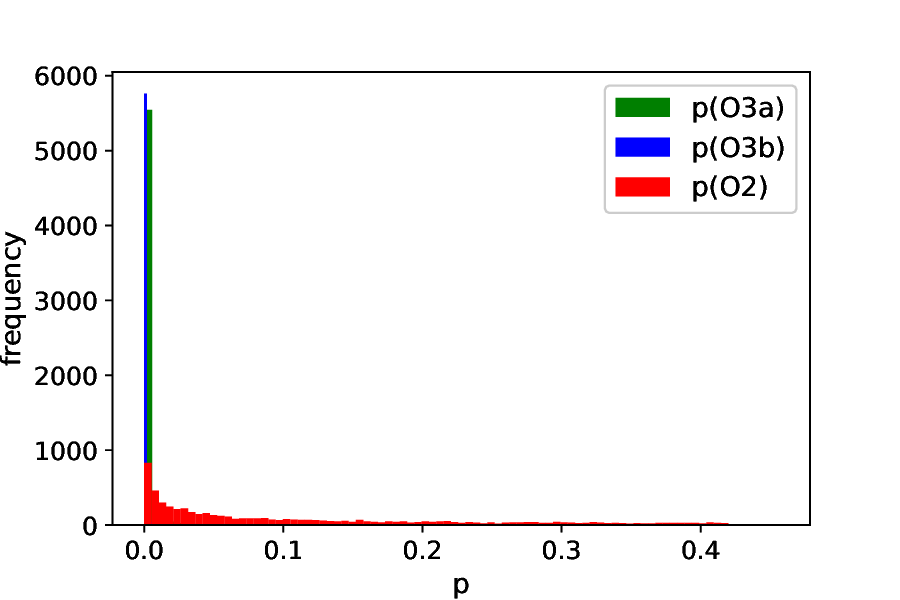} }}%
	\qquad
	\subfloat[\centering]{{\includegraphics[height = 4cm, width = 6cm]{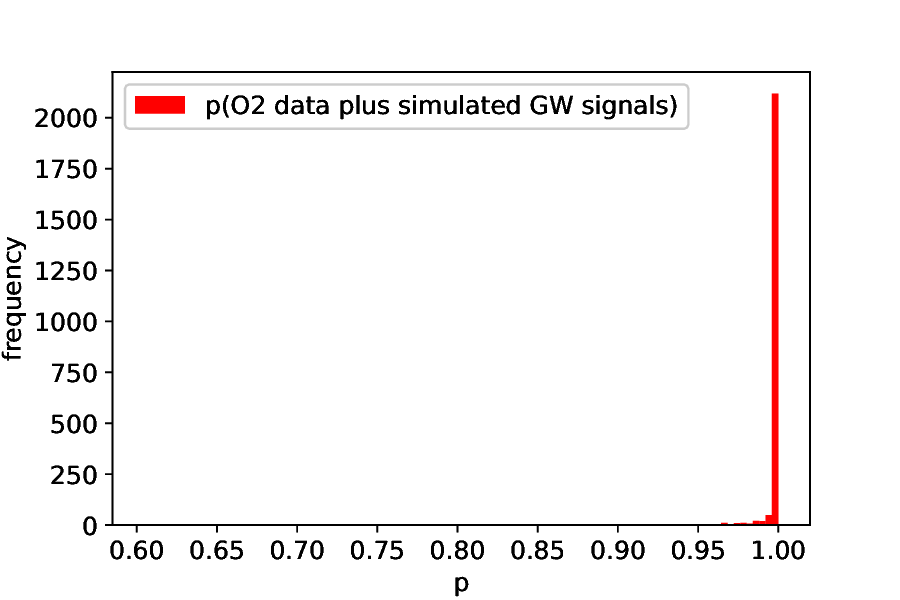} }}%
	\qquad
	\subfloat[\centering]{{\includegraphics[height = 4cm, width = 6cm]{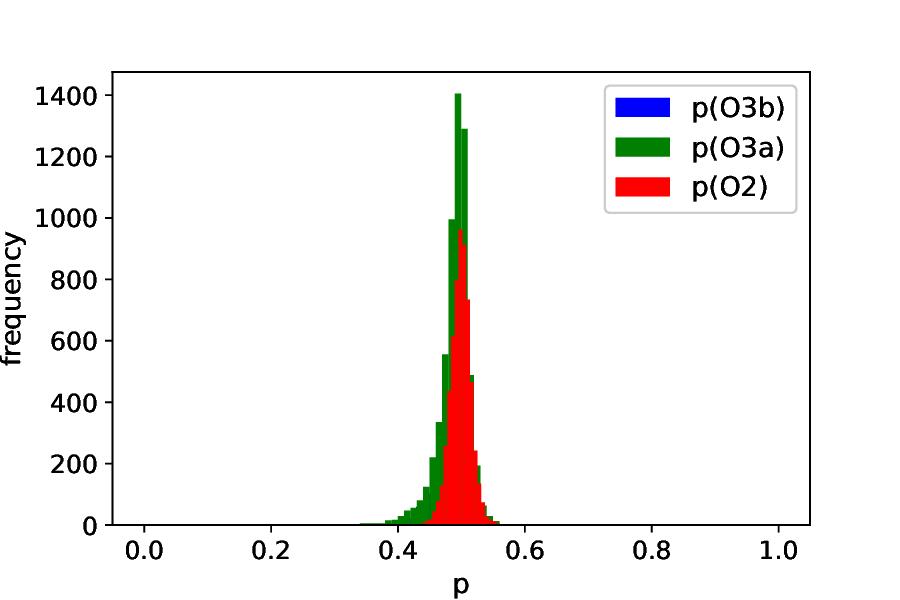} }}%
	\caption{Figures (a), (b) and (d) show the distribution of the confidence of containing typical targeted RGWs in the O2, O3a and O3b data for the sets 1, 2 and 3 in Table \ref{tab2}, respectively. The red, green and blue lines represent the distribution of the confidence of containing RGWs in the 6000 samples from the O2, O3a and O3b data, respectively. Fig (c) illustrates the distribution of the confidence of containing simulated RGWs in the O2 data for the set 1 in Table \ref{tab2}. The x-axis represents the distribution of $p$-values; the y-axis represents the distribution frequency, i.e., we divided the distribution of $p$-value  of 6000 samples into 100 narrow intervals, and count the number of samples with $p$-value sitting in each interval.}%
	\label{figre}%
\end{figure*}

\begin{figure*}
	\centering
	\subfloat[\centering]{{\includegraphics[height = 4cm, width = 6cm]{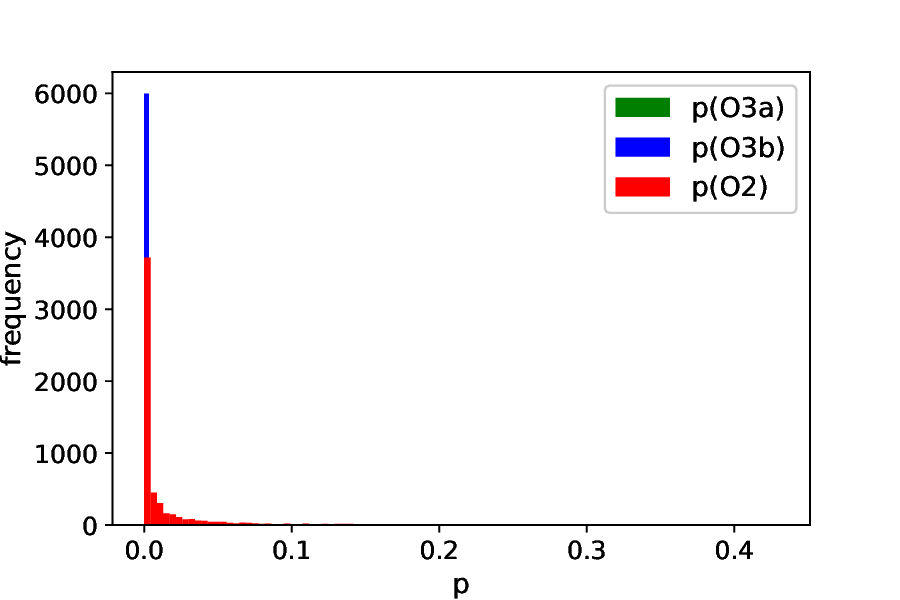} }}%
	\qquad
	\subfloat[\centering]{{\includegraphics[height = 4cm, width = 6cm]{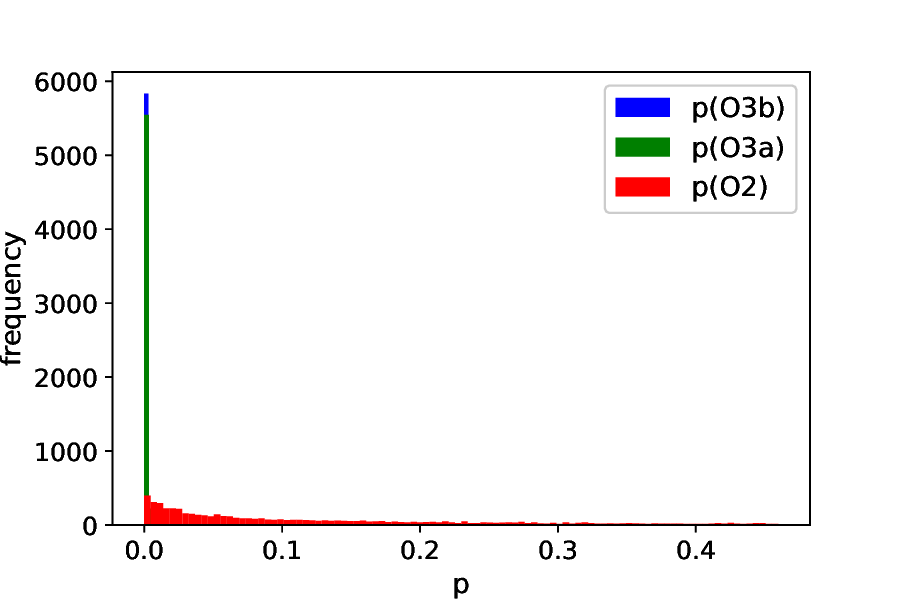} }}%
	\qquad
	\subfloat[\centering]{{\includegraphics[height = 4cm, width = 6cm]{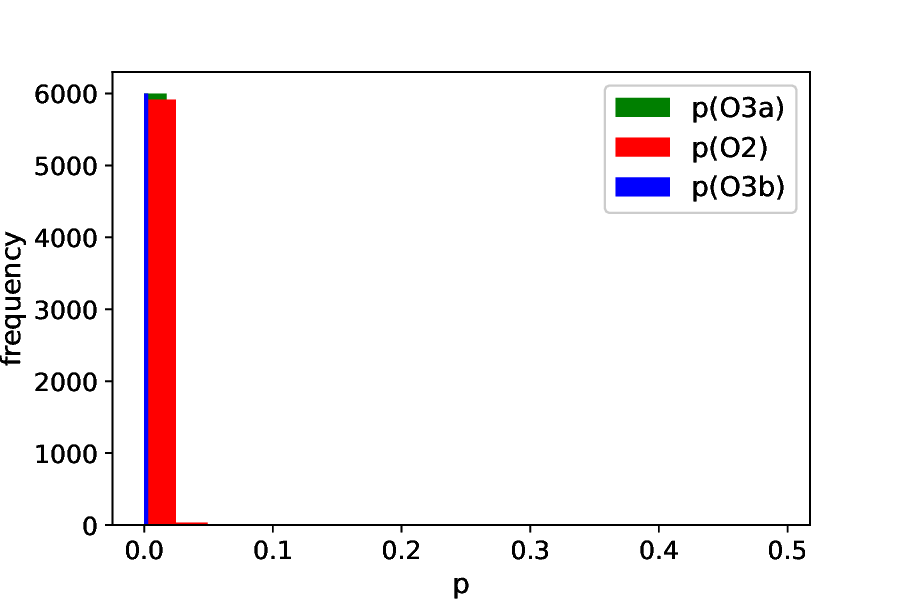} }}%
	\qquad
	\subfloat[\centering]{{\includegraphics[height = 4cm, width = 6cm]{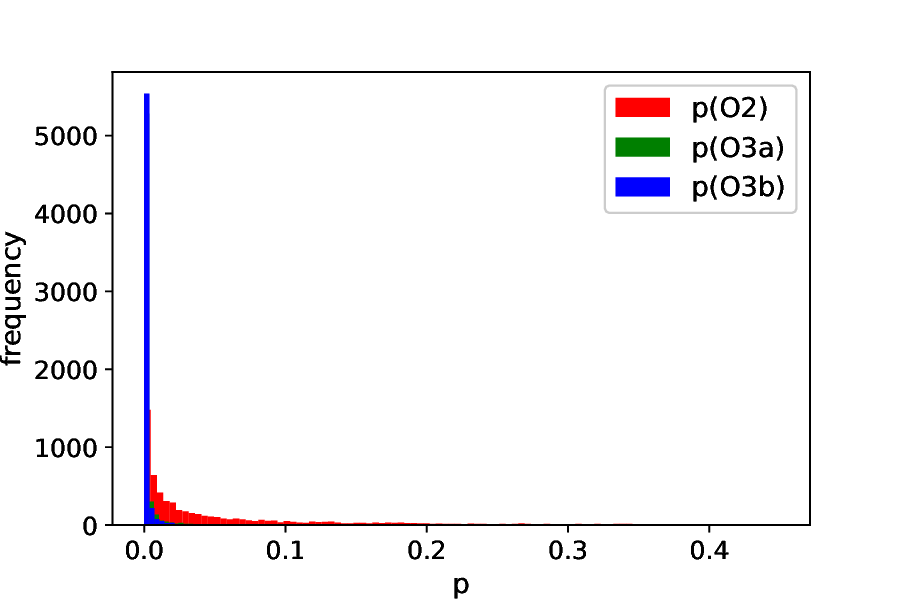} }}%
	\caption{Figures (a), (b), (c) and (d) demonstrate the distributions of the confidence of containing GWs from FOPT in the O2, O3a and O3b data  for the sets 1, 2, 4 and 5 in Table \ref{tab3}, respectively. The red, green and blue lines represent the distribution of the confidence of containing GWs from FOPT in the 6000 samples from the O2, O3a and O3b data, respectively.}%
	\label{figfo}%
\end{figure*}

From set 1 and 2 in Table \ref{tab2} and Fig. \ref{figre} (a) and (b), we can see that when the parameter spaces of the RGWinfl are [ $\beta\in(-1.8,-1.85)$, $\alpha\in(0.004,0.008)$ ] and [ $\beta\in(-1.85,-1.87)$, $\alpha\in(0.005,0.007)$ ], the magnitude of spectral energy density of RGWinfl is $\sim10^{-4}$ and $\sim10^{-5}$; the recognition accuracy  (the accuracy of the built CNN in determining whether the test data is mixed with the simulated GW signals, in addition to the real LIGO data) of the constructed CNN is around 0.99; the distributions of the confidence of containing RGWs in the 6000 samples from the O2, O3a and O3b data are less than 0.5, or, we found no evidence of RGW signals in the data obtained by the O2, O3a or O3b. From set 3 in Table \ref{tab2} and Fig. \ref{figre} (d), we can see that when the parameter space of the RGWs is [ $\beta\in(-1.88,-1.91)$, $\alpha\in(0.006,0.008)$ ], the magnitude of spectral energy density of RGWs is $\sim10^{-6}$;  the recognition accuracy of the built CNN are around 0.49; some of the 6000 sample data in O2, O3a and O3b contain RGW signals with confidence greater than 0.5, but we find that this is due to the false positives (the reason is about the invalidation of recognition for the GW signal parameters, will be explained later).

\indent From set 1 and 2 in Table \ref{tab3} and Fig. \ref{figfo} (a) and (b), we can see that for GWs from FOPT (sound wave case), where the parameter spaces of the GWs are [ $\beta/H_{\mathrm{pt}}\in(0.01,0.019)$, $\alpha\in(6.1,10)$, $T_{\mathrm{pt}}\in(7\times10^9,10^{10})$ Gev ] and  [ $\beta/H_{\mathrm{pt}}\in(0.02,0.16)$, $\alpha\in(1,10)$, $T_{\mathrm{pt}}\in(5\times10^9,10^{10})$ Gev ], the magnitude of spectral energy density of GWs is $\sim10^{-4}$ and $\sim10^{-5}$; the recognition accuracy of built CNN is around 0.99; the distributions of the confidence of containing GWs in the 6000 samples from the O2, O3a and O3b data are less than 0.5, or, we found no evidence of GW signals in the data obtained by the O2, O3a or O3b. From set 4 and 5 in Table \ref{tab3} and Fig. \ref{figfo} (c) and (d), when GWs originated from bubble collisions, where the parameter spaces of the GWs are [ $\beta/H_{\mathrm{pt}}\in(0.01,0.07)$, $\alpha\in(2,10)$, $T_{\mathrm{pt}}\in(10^9,10^{10})$ Gev] and [ $\beta/H_{\mathrm{pt}}\in(0.08,0.2)$, $\alpha\in(1,10)$, $T_{\mathrm{pt}}\in(5\times10^9,8\times10^{10})$ Gev], the magnitude of spectral energy density of GWs is $\sim10^{-4}$ and $\sim10^{-5}$; the recognition accuracy of established CNN is close to 0.99; the distributions of the confidence of containing GWs in the 6000 samples from the O2, O3a and O3b data are less than 0.5, and thus we found no evidence of GW signals in the data obtained by the O2, O3a or O3b.

\begin{figure}
	\centering
	\subfloat[\centering]{{\includegraphics[height = 3.7cm, width = 3.5cm]{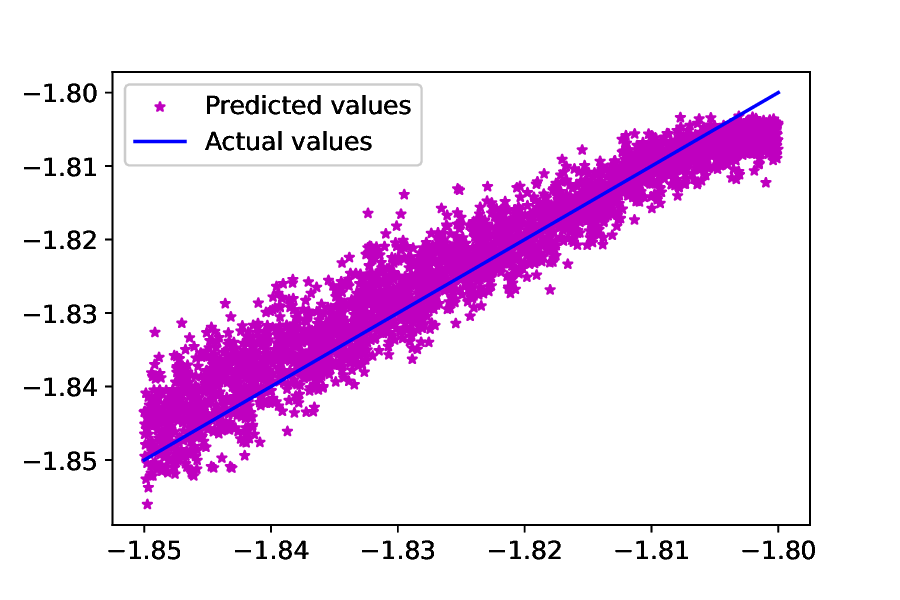} }}%
	\qquad
	\subfloat[\centering]{{\includegraphics[height = 3.7cm, width = 3.5cm]{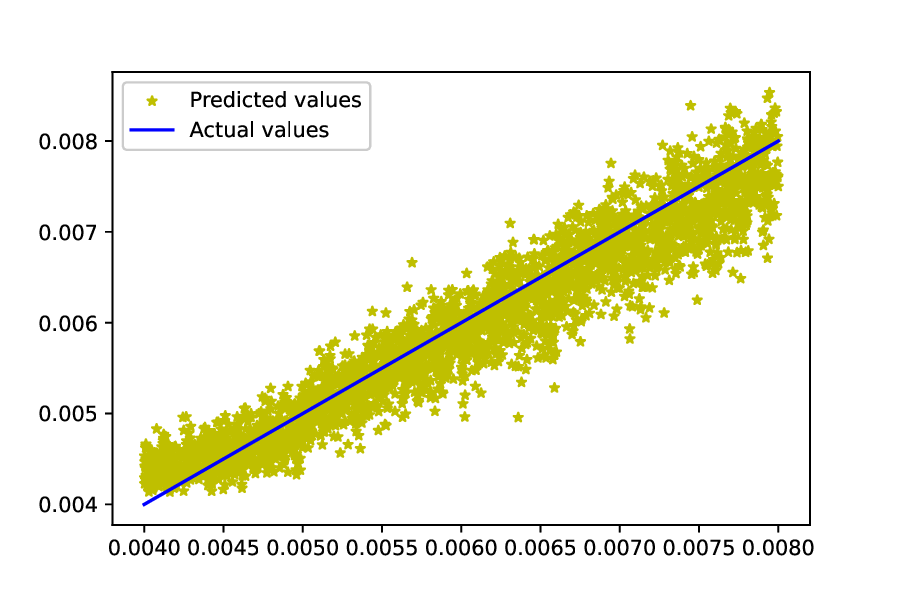} }}%
	\qquad
	\subfloat[\centering]{{\includegraphics[height = 3.7cm, width = 3.5cm]{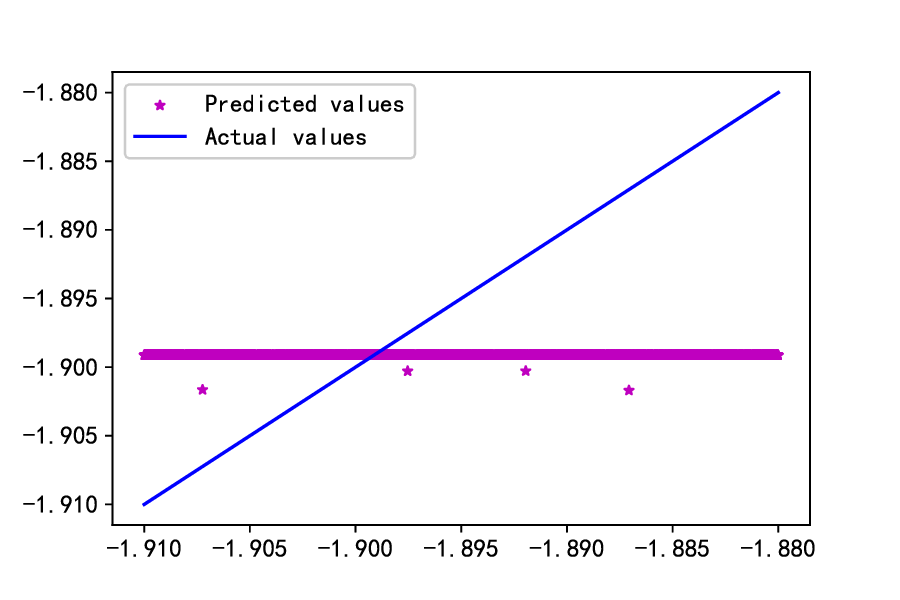} }}%
	\qquad
	\subfloat[\centering]{{\includegraphics[height = 3.7cm, width = 3.5cm]{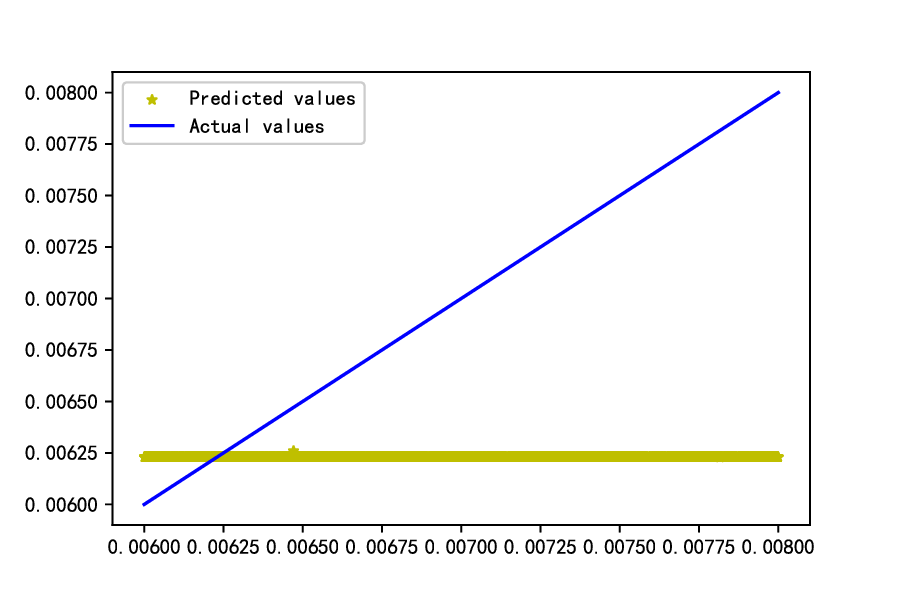} }}%
	\caption{Figures (a) and (b) express the results of simultaneous estimation of parameters of GWs $\beta$ and $\alpha$ for 10000 samples in set 1 in Table \ref{tab2}. Here $\beta\in(-1.8,-1.85)$, $\alpha\in(0.004,0.008)$. The residuals are 0.00071 and 0.00025, respectively. Figures (c) and (d) display the results of simultaneous estimation of $\beta$ and $\alpha$ for 10000 samples in set 3 of Table \ref{tab2}. Here $\beta\in(-1.88,-1.91)$, $\alpha\in(0.006,0.008)$. Importantly, the result shown in figures (c) and (d) indicate that for such order of magnitude of GW signals, the built CNN has no ability to correctly recognize the characteristics of the GW signals, or, the parameters cannot be properly estimated, so in such case the p-value given by the CNN is not reliable, and this is also the reason of the false positive in the subfigure (d) of Fig.3.}%
	\label{figesre}%
\end{figure}

\begin{figure*}
	\centering
	\subfloat[\centering]{{\includegraphics[height = 4cm, width = 4.2cm]{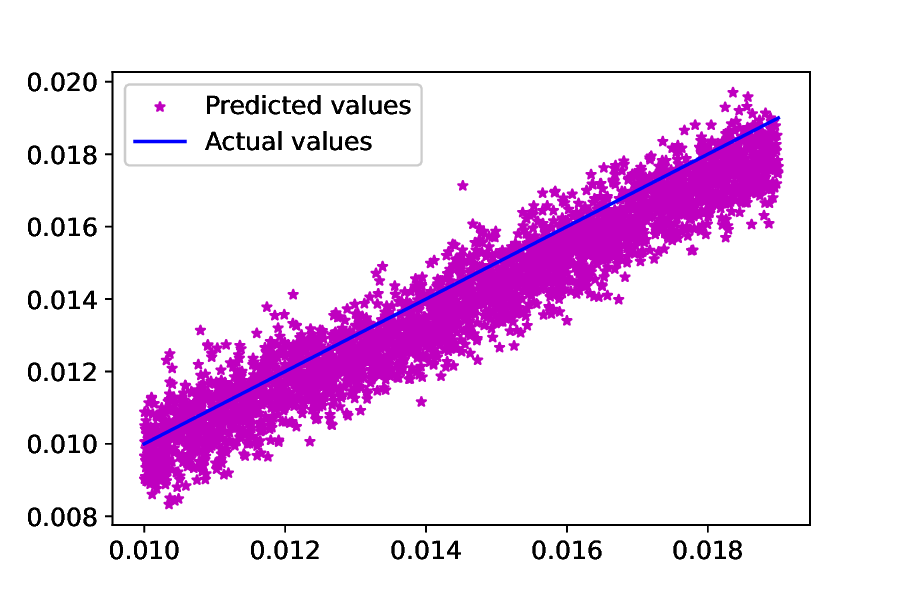} }}%
	\qquad
	\subfloat[\centering]{{\includegraphics[height = 4cm, width = 4.2cm]{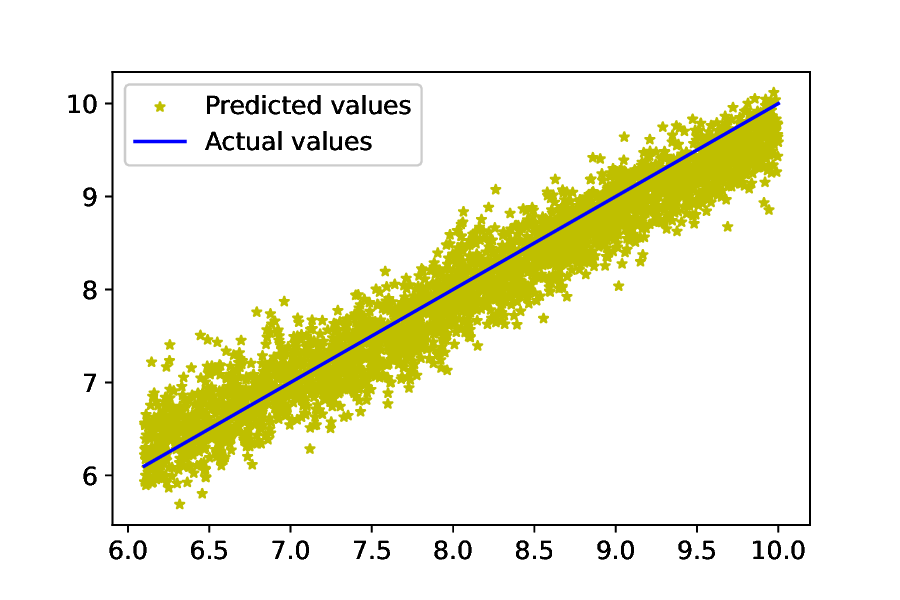} }}%
	\qquad
	\subfloat[\centering]{{\includegraphics[height = 4cm, width = 4.2cm]{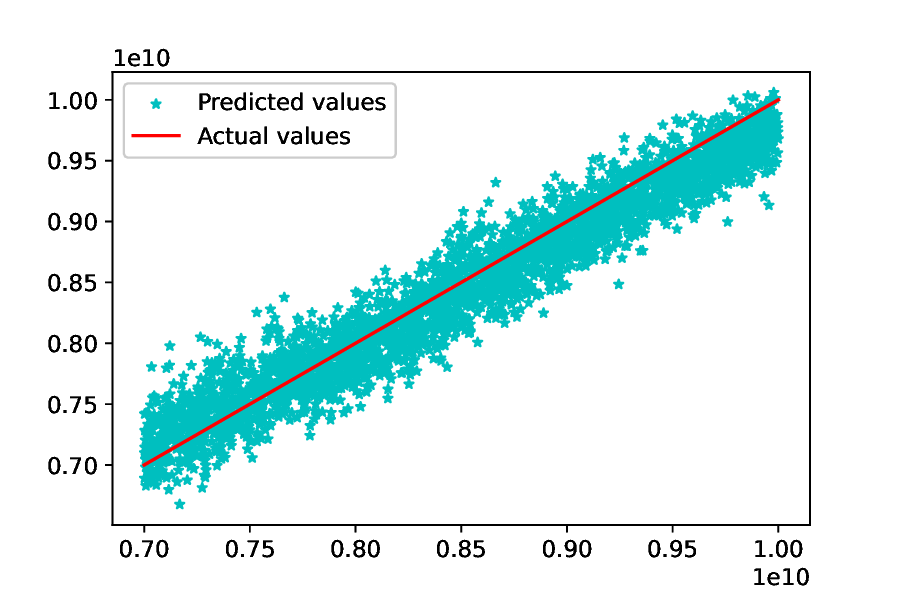} }}%
	\qquad
	\subfloat[\centering]{{\includegraphics[height = 4cm, width = 4.2cm]{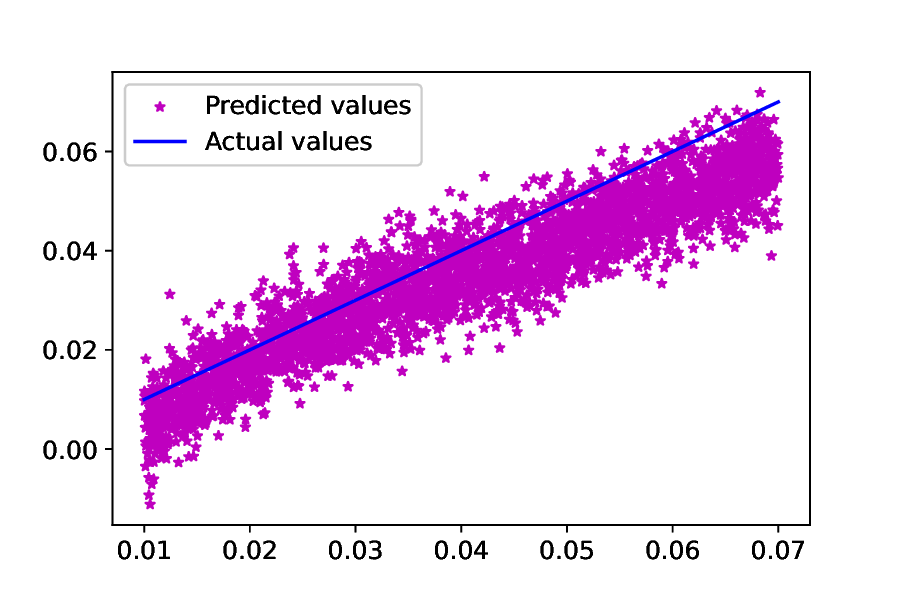} }}%
	\qquad
	\subfloat[\centering]{{\includegraphics[height = 4cm, width = 4.2cm]{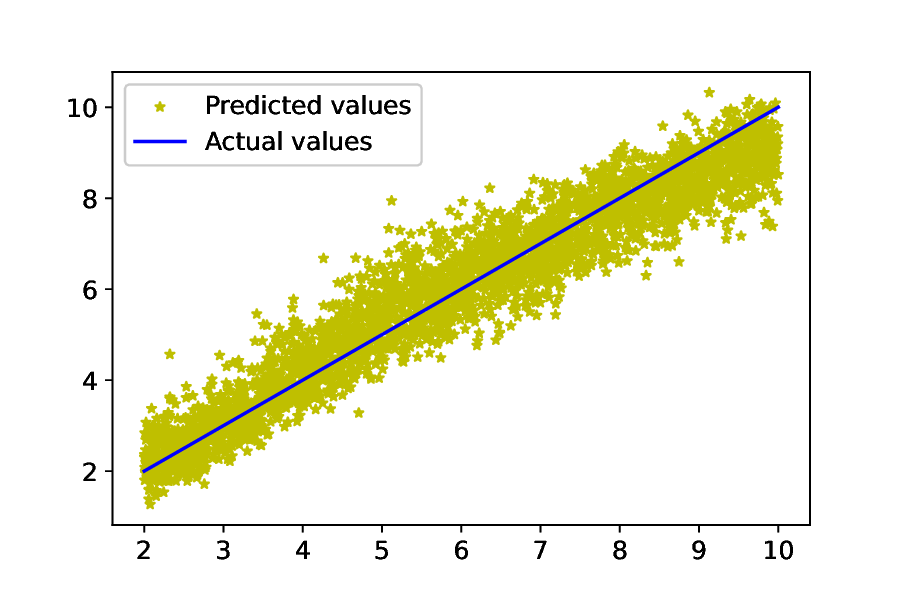} }}%
	\qquad
	\subfloat[\centering]{{\includegraphics[height = 4cm, width = 4.2cm]{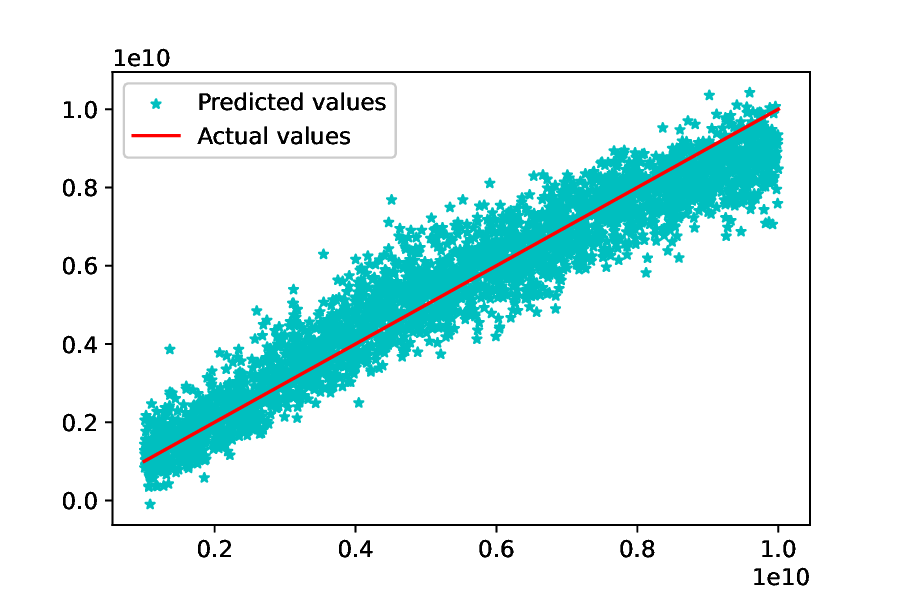} }}%
	\caption{Figures (a), (b) and (c) demonstrate the results of simultaneous  estimation of the parameters of GWs $\beta/H_{\mathrm{pt}}$, $\alpha$ and $T_{\mathrm{pt}}$ for 10000 samples in set 1 of Table \ref{tab3}. Here $\beta/H_{\mathrm{pt}}\in(0.01,0.019)$, $\alpha\in(6.	1,10)$, $T_{\mathrm{pt}}\in(7\times10^{9},10^{10})$Gev. The residuals are 0.000148, 0.0354 and 0.01026, respectively. Figures (c), (d) and (e) show the results of simultaneous  estimation of the parameters $\beta/H_{\mathrm{pt}}$, $\alpha$ and $T_{\mathrm{pt}}$ for 10000 samples in set 4 of Table \ref{tab3}. Here $\beta/H_{\mathrm{pt}}\in(0.01,0.07)$, $\alpha\in(2,10)$, $T_{\mathrm{pt}}\in(10^{9},10^{10})$Gev. The residuals are 0.00498, 0.01611 and 0.00148, respectively. The solid lines represent the actual values, and the dots represent the predicted values.}%
	\label{figesfo}%
\end{figure*}

\indent Next, we use constructed CNN (Table \ref{tab1}) to estimate the parameters of GW signals. The purpose of this task is to check whether the networks we establish   really have sufficiently acquired the ability to be able to recognize the features of the signal (i.e., be capable to correctly estimate the parameters of the GW signal). We consider three GW models of Eqs. (\ref{eqreg}), (\ref{eq4}) and (\ref{eq8}), respectively. For each type of GW model, within the set parameter spaces, we obtained a total of 100000 O2 data samples containing simulated GW signals. We use the mean squared error (MSE) loss function to evaluate the deviation between predicted and true values. In the final layer, we use the linear activation function to output parameter predicted values. From Figs. \ref{figesre} (a) (b) and \ref{figesfo}, we can see that our network model can recognize the characteristics of GW signals by correctly estimating 
corresponding multiple parameters of the GW signals.
As mentioned earlier, through reverse mapping, the selected sample points can cover the parameter regions corresponding to each order of magnitude of GW signals. Moreover, for dividing the training and testing sets, we also follow the principle of random distribution. Therefore, our network can reliably and robustly identify GW signals in specific order of magnitude among the parameter space. This ability is based on the confirmed capability of the network to recognize these concrete parameters, rather than relying on other arbitrary criteria or simply distinguishing between the presence or absence of any unknown signals. For example, as shown in Table 3 (line 5), the model exhibits a very high Recognition accuracy, so if any GW signal within the corresponding parameter space occurs, we have reason to believe that the model can detect such signal. However, if the result of test on the real LIGO data is null, it's reasonable  to believe that this null result is reliable, at least for the region of parameters we concern.
In contrary, the result shown in Figs. \ref{figesre} (c) and (d) indicate that for the order of magnitude ($\sim10^{-6}$) of RGWinfl signals, the built CNN has no ability to correctly recognize the characteristics of the RGWinfl signals, or, their parameters cannot be properly estimated, and this is also the reason of the false positive in the subfigure (d) of Fig. \ref{figre}.

\section{Conclusion and discussion}\label{sec4}

In this work, we try to search typical RGWs from early stage of universe as components of SGWB from the real LIGO data, which mainly involves typical RGWs from inflation and the GWs produced by the FOPT (mainly from sound waves and bubble collisions). By lengthy adjustment and adaptation process, we successfully establish effective and targeted deep learning neural network to calculate the likelihood that the real LIGO data contain above relic GW signals, and try to estimate their relevant parameters, or, to provide constraints on strengths of these RGWs. We find following results:

1. When the parameter spaces of the RGWinfl are [$\beta\in(-1.8,-1.85)$, $\alpha\in(0.004,0.008)$] and [$\beta\in(-1.85,-1.87)$, $\alpha\in(0.005,0.007)$]   (the corresponding magnitudes of spectral energy density of RGWinfl are $\sim10^{-4}$ and $\sim10^{-5}$), the distributions of the confidence of containing RGWinfl in the 6000 samples from the real LIGO data O2, O3a and O3b are less than 0.5, or, we found no evidence of RGWinfl signals in the data obtained by the O2, O3a or O3b. When the parameter spaces of the RGWinfl are [$\beta\in(-1.88,-1.91)$, $\alpha\in(0.006,0.008)$] (the corresponding magnitude of spectral energy density of RGWinfl is $\sim10^{-6}$), some of the 6000 sample data in O2, O3a and O3b seem contain RGW signals with confidence greater than 0.5; however, this result is due to the false positive, explained in Sect. \ref{sec3}, (also see Fig. \ref{figesre}), that, when the constructed CNN cannot correctly estimate the parameters of predicted GWs, the given p-value is not reliable or adoptable.

2. For GWs from FOPT (sound wave  case), with the parameter spaces of the GWs as  
[ $\beta/H_{\mathrm{pt}}\in(0.01,0.019)$, $\alpha\in(6.1,10)$, $T_{\mathrm{pt}}\in(7\times10^{9},10^{10})$ Gev] and [ $\beta/H_{\mathrm{pt}}\in(0.02,0.16)$, $\alpha\in(1,10)$, $T_{\mathrm{pt}}\in(5\times10^9,10^{10})$ Gev] (the corresponding magnitudes of spectral  energy density of GWs are $\sim10^{-4}$ and $\sim10^{-5}$), the distributions of the confidence of containing GWs in the 6000 samples from the real LIGO data O2, O3a and O3b are less than 0.5, or, we found no evidence of targeted GW signals in the data obtained by the O2, O3a or O3b. For case of GWs from bubble collisions, with the parameter spaces of the GWs as  [ $\beta/H_{\mathrm{pt}}\in(0.01,0.07)$, $\alpha\in(2,10)$, $T_{\mathrm{pt}}\in(10^9,10^{10})$ Gev] and [ $\beta/H_{\mathrm{pt}}\in(0.08,0.2)$, $\alpha\in(1,10)$, $T_{\mathrm{pt}}\in(5\times10^9,8\times10^{10})$ Gev] (the corresponding magnitudes of spectral energy density of GWs are $\sim10^{-4}$ and $\sim10^{-5}$), the distributions of the confidence of containing GWs in the 6000 samples from the real LIGO data O2, O3a and O3b are also less than 0.5, or, the same, we found no evidence of such GW signals in the data O2, O3a or O3b.\\ 

In fact, deep neural networks may result in misclassifications by omitting some features of signals, due to their limited generalization ability. However, in this paper, through reverse mapping, the selected sample points covered the corresponding parameter space for different order of magnitude,  thus would avoiding potential omissions. Furthermore, the neural network's binary classification decisions may also be affected by the presence of other false signals. However, through parameter estimation, we observe that the model possesses excellent parameter recognition capabilities. In other words, these abilities are based on accurate estimations of specific parameters, thereby excluding the possibility of other false signals being identified as true signals (as also described in the final part of section \ref{sec3}). For instance, when the GW parameters can be correctly recognized and estimated for cases with specific order of magnitude of GW strengths [Fig. 5(a), (b) and Fig. 6], once there is simulated GW signal mixed in the test data, the established CNN consistently gives very high `Recognition accuracy' and very high p-value [see sets 1, 2 of Table \ref{tab2}, and sets 1,2,4,5 of Table \ref{tab3}, and Fig. \ref{figre} (c)], so if the given p-value is low, it is believable that there is no targeted GW signal appearing; conversely, when the GW parameters cannot be correctly estimated for cases with some orders of GW strengths [e.g. Fig. \ref{figesre} (c) and (d)], even if the test data do contain the simulated GW signals, the CNN gives low Recognition accuracy and low p-value (e.g. set 3 of Table \ref{tab2}), so it turns unreliable, and thus for this situation, even if the CNN gives p-values greater than threshold of 0.5 after testing some data [e.g. Fig. \ref{figre} (d)], this still doesn't serve as an adoptable evidence for the existence of targeted GW signals. Therefore, with above consideration and operation, we could confirm the reliability of acquired results in this work.\\

Briefly, the results indicate no evidence of targeted GWs from 1)~inflation,  2)~sound waves, or 3)~bubble collisions, predicted by focused typical models, and for these 3 cases, the result provides upper limits of their  GW spectral energy densities, respectively as: 1)~$h^2\Omega_{GW}\sim10^{-5}$ [correspondingly in reversely mapped    2D parameter surface within rectangle of $\beta\in(-1.85,-1.87)$ $\times$ $\alpha\in(0.005,0.007)$],   and 2)~$h^2\Omega_{GW}\sim10^{-5}$ [in reversely mapped 3D parameter volume within cuboid of $\beta/H_{\mathrm{pt}}\in(0.02,0.16)$ $\times$ $\alpha\in(1,10)$ $\times$ $T_{\mathrm{pt}}\in(5\times10^{9},10^{10})$ Gev], and 3)~$h^2\Omega_{GW}\sim10^{-5}$ [in reversely mapped 3D parameter volume within cuboid of $\beta/H_{\mathrm{pt}}\in(0.08,0.2)$ $\times$ $\alpha\in(1,10)$ $\times$ $T_{\mathrm{pt}}\in(5\times10^{9},8\times10^{10})$ Gev].
Although only null results and upper limits are acquired here, above methods and neural networks we develop to search RGWs from LIGO data could be effective and reliable, and it can be applied not only for currently captured data but also for upcoming data by O4 running or other observational datasets, to establish an available scheme for exploring possible RGWs from very early stage of the cosmos or to provide constraints on relevant cosmological theories.

\begin{acknowledgments}
This work was supported in part by the National Natural Science Foundation of China under Grant No.11605015, No.12347101 and No.12147102, the Natural Science Foundation of Chongqing, China (Grant No.: cstc2020jcyj-msxmX0944), and the Research Funds for the Central Universities (2022CDJXY-002).
\end{acknowledgments}

\bibliography{Likelihoodbib}

\end{document}